\documentclass[article,superscriptaddress,twocolumn]{revtex4-1}

\usepackage{amssymb}
\usepackage{amsfonts}
\usepackage{amsmath}
\usepackage{graphicx}
\usepackage{dcolumn}
\usepackage{bm}
\usepackage{xcolor}
\usepackage{float}
\usepackage{textgreek}
\usepackage{upgreek}

\usepackage[colorlinks]{hyperref}
\hypersetup{
                pdfstartview={FitH},
                linkcolor=blue,
                citecolor=blue,
                filecolor=blue,
                urlcolor=blue
}

\begin{document}

\begin{sloppypar}

\title{Spin-orbitronics at a topological insulator-semiconductor interface}

\author{T. Guillet}
\thanks{These two authors equally contributed to the present work.}
\affiliation{Univ. Grenoble Alpes, CEA, CNRS, Grenoble INP, IRIG-Spintec, 38000 Grenoble, France}

\author{C. Zucchetti}
\thanks{These two authors equally contributed to the present work.}
\affiliation{LNESS-Dipartimento di Fisica, Politecnico di Milano, Piazza Leonardo da Vinci 32, 20133 Milano, Italy}

\author{A. Marchionni}
\affiliation{LNESS-Dipartimento di Fisica, Politecnico di Milano, Piazza Leonardo da Vinci 32, 20133 Milano, Italy}

\author{A. Hallal}
\affiliation{Univ. Grenoble Alpes, CEA, CNRS, Grenoble INP, IRIG-Spintec, 38000 Grenoble, France}

\author{P. Biagioni}
\affiliation{LNESS-Dipartimento di Fisica, Politecnico di Milano, Piazza Leonardo da Vinci 32, 20133 Milano, Italy}

\author{C. Vergnaud}
\affiliation{Univ. Grenoble Alpes, CEA, CNRS, Grenoble INP, IRIG-Spintec, 38000 Grenoble, France}

\author{A. Marty}
\affiliation{Univ. Grenoble Alpes, CEA, CNRS, Grenoble INP, IRIG-Spintec, 38000 Grenoble, France}

\author{M. Finazzi}
\affiliation{LNESS-Dipartimento di Fisica, Politecnico di Milano, Piazza Leonardo da Vinci 32, 20133 Milano, Italy}

\author{F. Ciccacci}
\affiliation{LNESS-Dipartimento di Fisica, Politecnico di Milano, Piazza Leonardo da Vinci 32, 20133 Milano, Italy}

\author{M. Chshiev}
\affiliation{Univ. Grenoble Alpes, CEA, CNRS, Grenoble INP, IRIG-Spintec, 38000 Grenoble, France}

\author{F. Bottegoni}
\affiliation{LNESS-Dipartimento di Fisica, Politecnico di Milano, Piazza Leonardo da Vinci 32, 20133 Milano, Italy}

\author{M. Jamet}
\affiliation{Univ. Grenoble Alpes, CEA, CNRS, Grenoble INP, IRIG-Spintec, 38000 Grenoble, France}

\date{\today}


\begin{abstract}

Topological insulators (TIs) hold great promises for new spin-related phenomena and applications thanks to the spin texture of their surface states. However, a versatile platform allowing for the exploitation of these assets is still lacking due to the difficult integration of these materials with the mainstream Si-based technology. Here, we exploit germanium as a substrate for the growth of Bi$_2$Se$_3$, a prototypical TI. We probe the spin properties of the Bi$_2$Se$_3$/Ge pristine interface by investigating the spin-to-charge conversion taking place in the interface states by means of a non-local detection method. The spin population is generated by optical orientation in Ge, and diffuses towards the Bi$_2$Se$_3$ which acts as a spin detector. We compare the spin-to-charge conversion in Bi$_2$Se$_3$/Ge with the one taking place in Pt in the same experimental conditions. Notably, the sign of the spin-to-charge conversion given by the TI detector is reversed compared to the Pt one, while the efficiency is comparable. By exploiting first-principles calculations, we ascribe the sign reversal to the hybridization of the topological surface states of Bi$_2$Se$_3$ with the Ge bands. These results pave the way for the implementation of highly efficient spin detection in TI-based architectures compatible with semiconductor-based platforms. 

\end{abstract}

\maketitle

In the past decade, the discovery of topological insulators (TIs) has promised a breakthrough in the efficiency of  spin-charge interconversion phenomena. Indeed, TIs are known to host topologically-protected surface states (TSS) leading to spin-momentum locking \cite{Hasan2010}. This has been experimentally verified by means of photoemission measurements \cite{Hsieh2009,Zhang2009}, scanning tunneling microscopy, and magnetotransport experiments \cite{Kim2011,Liu2012,Lang2012}. In particular, spin-momentum locking in TSS leads to the conversion of a charge current into a spin current, a phenomenon that is commonly addressed as the Rashba-Edelstein effect (REE), while the reverse process is referred to as the inverse Rashba-Edelstein effect (IREE) \cite{Edelstein1990}. In these systems, the leading parameters are the spin-charge interconversion efficiencies: ${q_{\textup{REE}}=j_{\textup{s}}^{3\textup{D}}/j_{\textup{c}}^{2\textup{D}}}$ for the REE and ${\lambda_{\textup{IREE}}=j_{\textup{c}}^{2\textup{D}}/j_{\textup{s}}^{3\textup{D}}}$ for the IREE. However, an experimental estimation based on spin pumping-ferromagnetic resonance (FMR) or spin torque-FMR \cite{Jamali2015,Shi2018,Wang2016} is questionable, since TIs are known to chemically react when they are in contact with a ferromagnetic film \cite{Walsh2017,Ferfolja2018}. Therefore, a non-local architecture where the source of the spin current and the TI are well separated would represent a reliable route to avoid the aforementioned issue.

In this letter, we use germanium as a platform for such non-local spin-to-charge conversion (SCC) measurements. The spin population is generated by optical spin orientation in Ge, and diffuses as a spin current towards the Bi$_2$Se$_3$, which acts as the spin detector. In this way, we totally avoid any ferromagnetic material to generate the spin current. 

Bi$_2$Se$_3$ is grown by van der Waals epitaxy on Ge(111) \cite{Guillet2018}. We probe SCC at the Bi$_{2}$Se$_{3}$/Ge interface kept at room temperature and compare the experimental results with those obtained from a Pt/Ge junction in the same experimental conditions. We then estimate the SCC efficiency in Bi$_2$Se$_3$  and Pt, which are found to be of the same order of magnitude. We also evaluate the conversion efficiency of the IREE in the junction to be ${\lambda_{\textup{IREE}}\approx -30~\text{pm}}$. Notably, the sign of the SCC is opposite in the two cases. To understand this sign reversal, we employ first principles calculations and demonstrate the existence of Rashba states at the Bi$_2$Se$_3$/Ge interface as a result of strong interfacial hyridization. We find that these states exhibit an opposite spin chirality compared to the one of TSS in bulk-terminated Bi$_2$Se$_3$.

\begin{figure}[t]
\begin{center}
\includegraphics[width=0.45\textwidth]{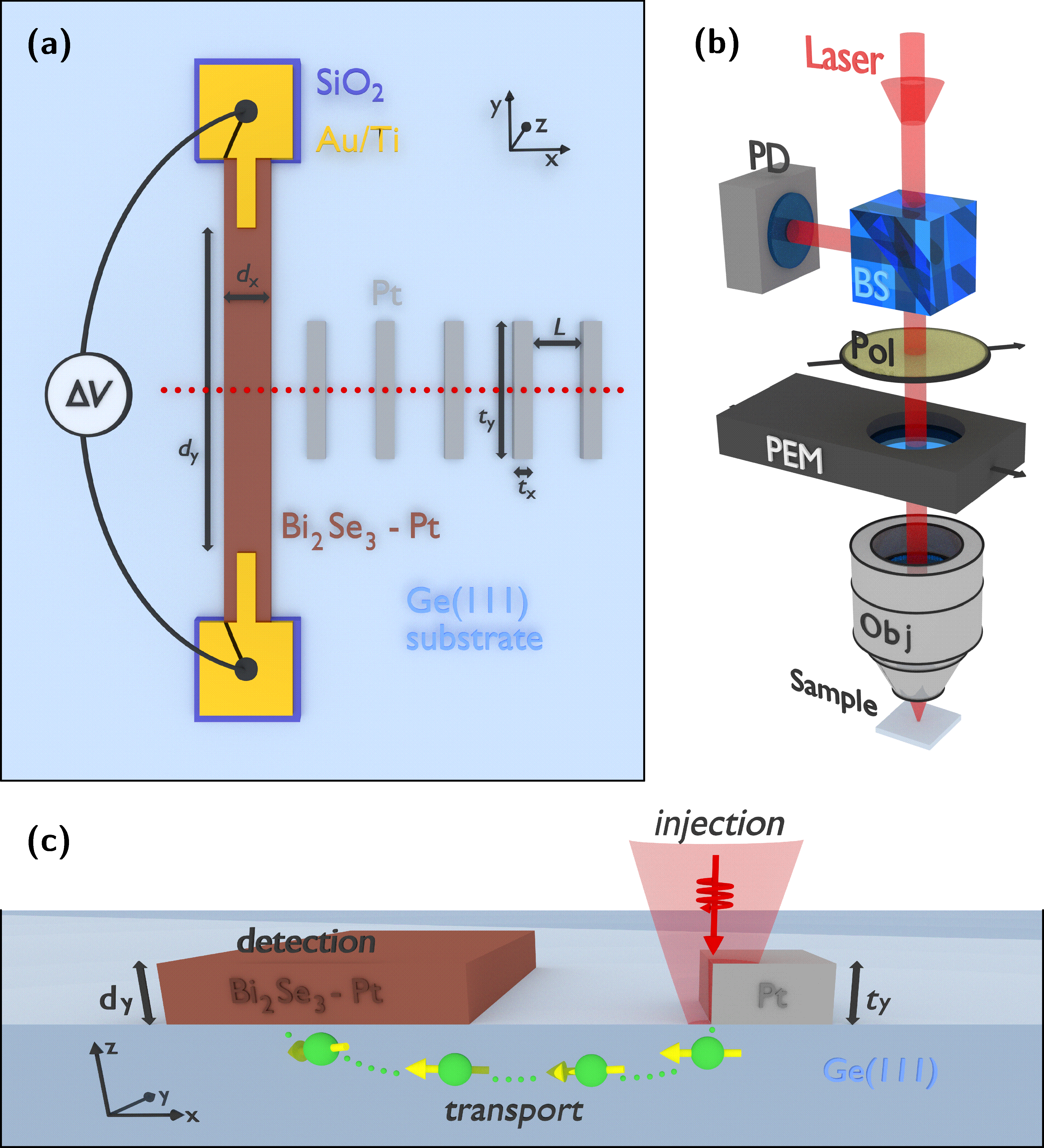}
\end{center}
\caption{(Color online) (a) Sample layout. (b) Confocal microscopy setup. (c) Spin generation and diffusion in Ge(111)} \label{fig1}
\end{figure}

The investigated samples are sketched in Fig.~\ref{fig1}\textcolor{blue}{(a)}. As a substrate, we use a ${2~\text{\textmu m-thick}}$ \textit{n}-doped Ge(111) layer (doping concentration ${N_{\textup{d}}=9\times10^{16}~\text{cm}^{-3}}$) epitaxially grown on semi-insulating Si. On this substrate we lithographically define a set of ${20\times2~\text{\textmu m}^2}$ Pt stripes with a thickness of ${15~\text{nm}}$, which are exploited for optical spin injection \cite{Bottegoni2014,Zucchetti2017}. The detection of spin-polarized electrons is performed in a ${75\times20~\text{\textmu m}^2}$ bar of either hexagonal single crystalline Bi$_2$Se$_3$ (${10~\text{nm-thick}}$) or Pt (${15~\text{nm-thick}}$). In both cases, the detection bar is electrically contacted by two Ti/Au pads. To prevent direct absorption of spins from these contacts, a 80 nm-thick SiO$_2$ layer is inserted between the electrodes and Ge. The measurements have been performed at room temperature using a confocal microscope, shown in Fig.~\ref{fig1}\textcolor{blue}{(b)}. The energy of the photons is tuned to the direct Ge gap (${\hbar\omega=0.8~\text{eV}}$) and the circular polarization of the light is modulated by a photoelastic modulator (PEM) at 50 kHz. The light is then focused on the sample by an objective with a 0.7 numerical aperture, yielding a spot of full-size diameter on the sample of about ${3~\text{\textmu m}}$. The voltage drop $\Delta V$ between the two Ti/Au pads is then obtained by demodulating with a lock-in amplifier the signal acquired under open-circuit conditions between the electrodes [see Fig.~\ref{fig1}\textcolor{blue}{(a)}] while the focused light beam raster scans the sample surface.

In Fig.~\ref{fig1}\textcolor{blue}{(c)}, we report the working principle of the non-local spin-injection/detection scheme. When the circularly-polarized light beam impinges on the edge of a Pt stripe, a highly-localized spin population polarized along the $x$-axis is generated, through the optical spin orientation technique \cite{Meier1984}, in the Ge layer underneath the Pt edge \cite{Bottegoni2014}. If the light impinges on the opposite edge of the Pt stripe, the direction of the spin-polarization parallel to the $x$-axis is reversed, as previously demonstrated in Ref.~\onlinecite{Bottegoni2014}. 
After generation, the spin-polarized electrons diffuse in the Ge substrate toward the detection bar, where SCC occurs and the spin current is converted in a charge accumulation that creates an electric field directed along $y$. The latter is eventually detected as a voltage drop $\Delta V$ between the electrodes [see Fig.~\ref{fig1}\textcolor{blue}{(a)}]. The spin-to-charge conversion is caused by the inverse spin-Hall effect (ISHE) \cite{Dyakonov1971} in the case of detection with Pt, and by the IREE \cite{Edelstein1990} in the case of Bi$_2$Se$_3$. In both cases, the geometry of spin-to-charge conversion imposes $\Delta V$ to be sensitive only to an electron spin polarization directed along the $x$-axis \cite{Ando2010,Zucchetti2017}.

\begin{figure}[t]
\begin{center}
\includegraphics[width=0.48\textwidth]{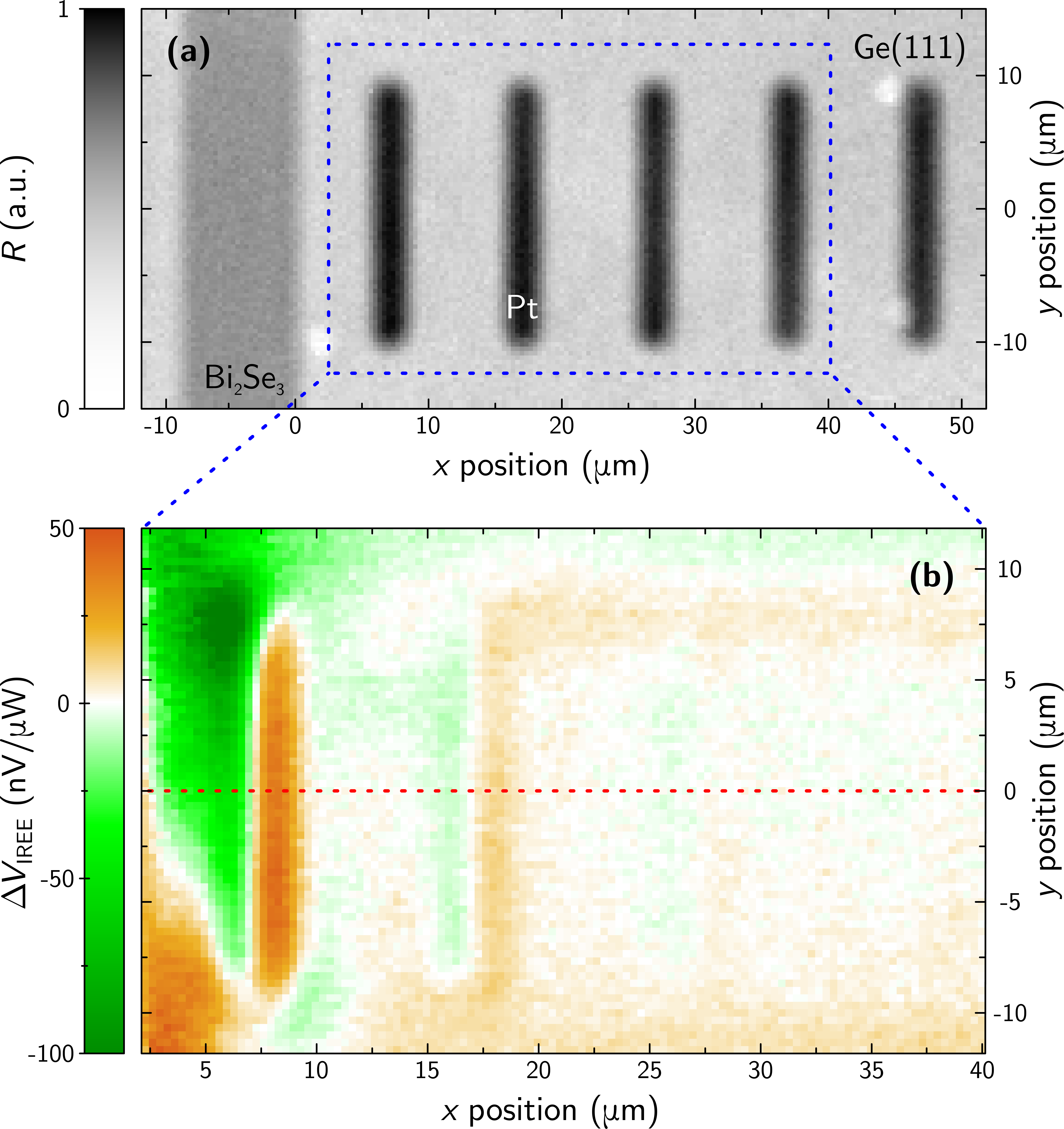}
\end{center}
\caption{(Color online) Reflectivity (a) and IREE map (b) of the Bi$_2$Se$_3$/Ge(111) junction acquired for an incident power ${W=18~\text{\textmu W}}$ at ${\hbar\omega=0.8~\text{eV}}$. The IREE map is collected over the area identified by the dashed blue rectangle in the upper panel.} \label{fig2}
\end{figure}

\begin{figure}[t]
\begin{center}
\includegraphics[width=0.48\textwidth]{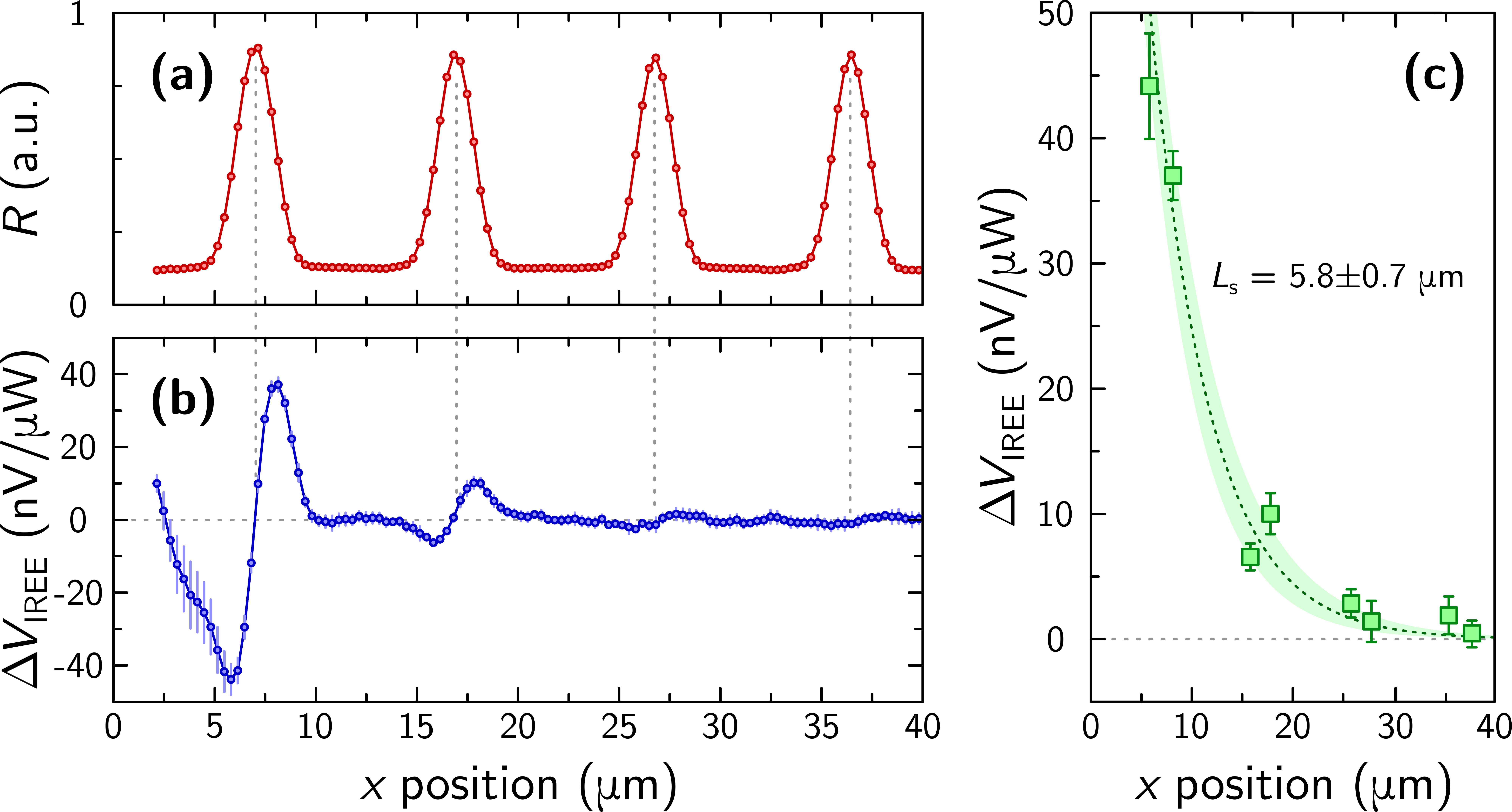}
\end{center}
\caption{(Color online) Reflectivity (a) and IREE (b) profiles signal along the $x$-axis of the sample. (c) Absolute value of $\Delta V_\mathrm{IREE}$ as a function of $x$ at correspondence with the right edge of each Pt microstructure.} \label{fig3}
\end{figure}

We first show the results obtained on the sample with the Bi$_2$Se$_3$ detector. The reflectivity and the electrical maps are shown in Fig.~\ref{fig2}\textcolor{blue}{(a)} and \textcolor{blue}{(b)}, respectively. The electrical map is normalized to the impinging laser power $W$. As expected for a spin-related signal, by illuminating at opposite edges the Pt stripes used for spin injection, the sign of the electric signal is reversed. This can be better visualized in Fig.~\ref{fig3} that shows the profiles, integrated along the $y$-axis, of the reflectivity [panel \textcolor{blue}{(a)}] and electrical maps [panel \textcolor{blue}{(b)}]. 
From the latter, we also observe the decrease of the absolute value of the signal when the generation point (i.e., the edge of the illuminated Pt stripe) moves away from the Bi$_2$Se$_3$ bar. This signal decay is related to the spin depolarization from the generation to the detection point, which is larger for longer paths \cite{Zucchetti2017}. We can see better this behavior in Fig.~\ref{fig3}$\,$\textcolor{blue}{(c)} that reports the absolute value of the electrical signal measured at each Pt edge as a function of the distance $x$ from the position of the Bi$_2$Se$_3$ detector. Exploiting a simple unidimensional spin-diffusion model ${\Delta V_\mathrm{IREE}\propto e^{-x/L_\mathrm{s}}}$ \cite{Zucchetti2017} to fit the experimental data, we estimate ${L_\mathrm{s}=5.8\pm0.7~\text{\textmu m}}$, slightly shorter than the one reported for Ge(001) substrates with similar doping \cite{Zucchetti2019}. 

\begin{figure}[t]
\begin{center}
\includegraphics[width=0.48\textwidth]{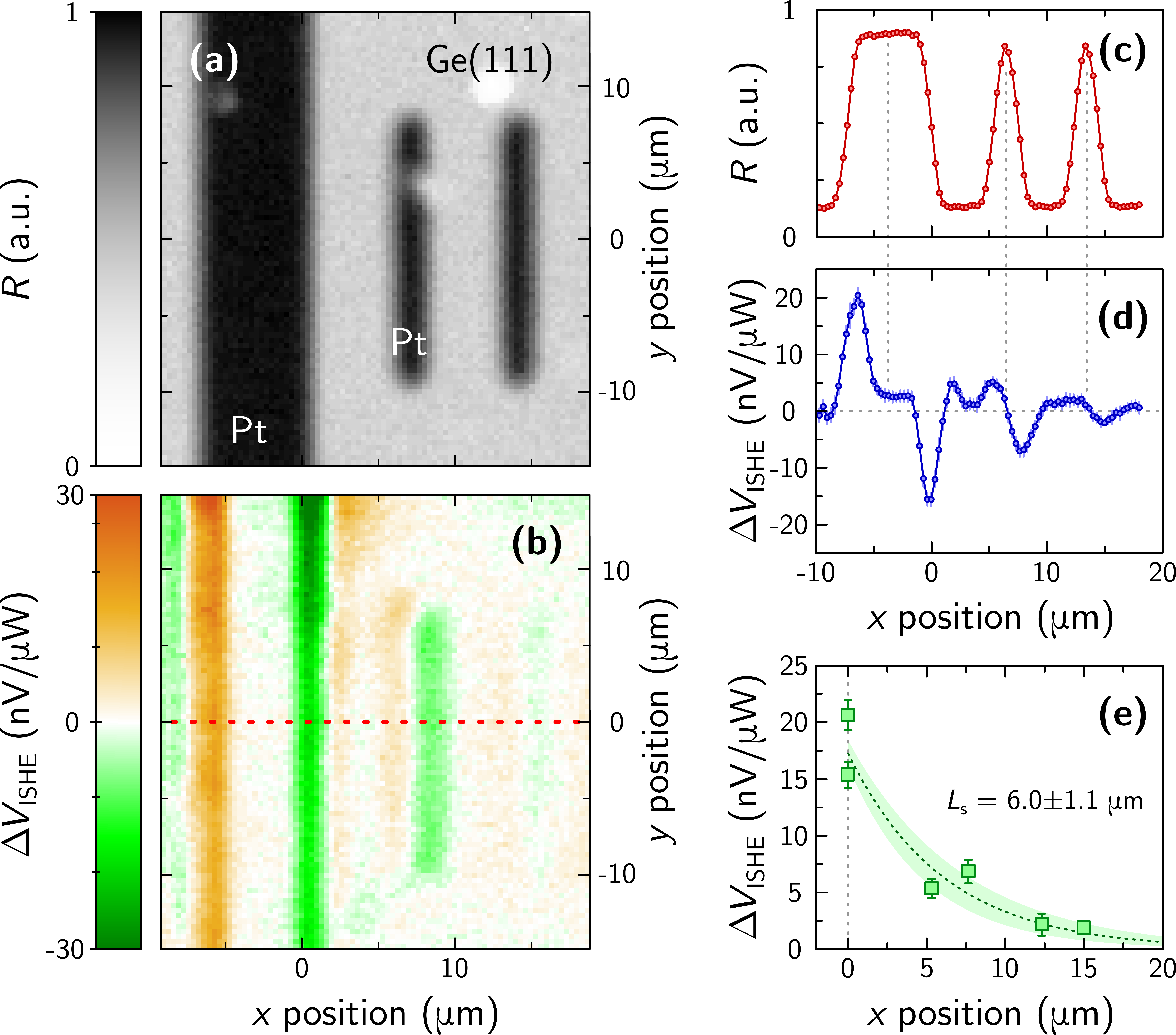}
\end{center}
\caption{(Color online) Reflectivity (a) and ISHE map (b) of the Pt/Ge(111) sample acquired for ${W=15~\text{\textmu W}}$ at ${\hbar\omega=0.8~\text{eV}}$. Reflectivity (c) and ISHE (d) profiles along the $x$-axis of the sample. (e) Amplitude of the ISHE signal at correspondence with the right edge of each Pt microstructure.} \label{fig4}
\end{figure}

The same analysis, summarized in Fig.~\ref{fig4}, has been performed for the sample with the Pt detector. Panels \textcolor{blue}{(a,b)} show the reflectivity and the normalized ISHE map of the sample, respectively, while in panels \textcolor{blue}{(c,d)} we report the profiles along the $x$ axis of the two maps. In panel \textcolor{blue}{(e)}, we plot the dependence of the absolute value of the ISHE signal acquired at correspondence with the Pt edges as a function of the distance from the spin detector. In this case, the fitting with the spin-diffusion model presented above yields ${L_\mathrm{s}=6.0\pm1.1~\text{\textmu m}}$, which perfectly matches the value obtained with the Bi$_2$Se$_3$ sample.

Since the spin-injection and transport mechanisms are the same for the two samples, it is possible to quantitatively compare the results obtained with Bi$_2$Se$_3$ and Pt detectors. First, from Fig.~\ref{fig3}\textcolor{blue}{(a,b)}, we observe that the Bi$_2$Se$_3$ detection provides a negative (positive) voltage drop when the focused light beam illuminates the left (right) edge of the Pt injection stripes. Conversely, when SCC is performed via the ISHE in Pt, the signal is positive (negative) at the left (right) edge of the injector microstructures [see Fig.~\ref{fig4}\textcolor{blue}{(c,d)}]. Hence, the sign of the spin-to-charge conversion in Bi$_2$Se$_3$/Ge is found to be opposite to that in Pt. Previous experiments were performed to characterize the spin-to-charge conversion in Bi$_2$Se$_3$ thin films \cite{Jamali2015,Shi2018,Wang2016} and, at variance with our result, the conversion parameter was always measured with the same sign as ISHE in Pt, which we arbitrarily define as \textquotedblleft positive\textquotedblright. 
Although the SCC measurements in Refs.~\citenum{Jamali2015,Shi2018,Wang2016} were carried out with the TI in direct contact with a ferromagnet, this positive sign is also expected from photoemission spectroscopy \cite{Hsieh2009,Neupane2014} and electrical spin detection \cite{Li2014,Lee2015}. Hence, as further discussed in the following, our experimental results suggest that the spin-split states at the Bi$_2$Se$_3$/Ge(111) interface display a substantially different SCC behavior compared to the ones of a freestanding Bi$_2$Se$_3$ surface.

Beyond this sign reversal, the comparison of Fig.~\ref{fig3}\textcolor{blue}{(c)} and Fig.~\ref{fig4}\textcolor{blue}{(e)} allows one to estimate the relative spin detection efficiency of Bi$_2$Se$_3$/Ge and Pt. With the light beam focused on the first Pt stripe (${x=x_0\approx6~\text{\textmu m}}$), we measure ${\Delta V_{\textup{IREE}}/W\approx40~\text{nV/\textmu W}}$ for Bi$_2$Se$_3$/Ge and ${\Delta V_{\textup{ISHE}}/W\approx7~\text{nV/\textmu W}}$ for Pt. Since the two samples only differ by the spin detector, we conclude that the overall efficiency for spin detection with Bi$_2$Se$_3$/Ge is a factor 5 larger than Pt. The insulating character of bulk TIs indeed produces higher voltage drops compared to a metal like Pt for the same charge current.

The macroscopic spin-to-charge conversion parameter is $\gamma=i_{\textup{c}}/i_{\textup{s}}$, being $i_\mathrm{s}$ the spin current entering the detector and $i_\mathrm{c}$ the equivalent charge current across the detection bar, defined as the ratio between the open circuit ISHE or IREE signal $\Delta V$ and the detector resistance $R$, $i_\mathrm{c}=\Delta V/R$. If we assume the same value of $i_\textup{s}$ for the two samples (due to equal spin injection and transport mechanisms), the relative spin-to-charge conversion efficiency of the materials is ${\gamma_{\textup{BiSe}}/\gamma_{\textup{Pt}}=i_{\textup{c,BiSe}}/i_{\textup{c,Pt}}}$. Considering the $\Delta V$ values recorded at ${x=x_0\approx6~\text{\textmu m}}$ and being ${R_{\textup{BiSe}}\approx10~\text{k}\Omega}$ and ${R_{\textup{Pt}}\approx500~\Omega}$ the detector resistance, as measured by a four-probe technique, we obtain ${\gamma_{\textup{Pt}}\approx-3.5\,\gamma_{\textup{BiSe}}}$. The absolute determination of $\gamma$ requires the knowledge of $i_{\textup{s}}$. To estimate its value, we start from the spin current excited at the generation time:
\begin{equation}
i_\mathrm{s,0}=\frac{T\,W}{\hbar\omega}\,P\,\eta_{\textup{g}},
\end{equation}
where ${T\,W/(\hbar\omega)}$ is the photon absorption rate (${T\approx0.6}$ is the transmittance of Ge at ${\hbar\omega=0.8~\textup{eV}}$ and $W$ the impinging light power), ${\eta_{\textup{g}}}$ the fraction of absorbed photons with a projection of the angular momentum along the $x$-axis and $P=50\,\%$ \cite{Meier1984} is the ratio between spin-polarized photogenerated electrons and absorbed photons. Exploiting finite-difference time-domain three-dimensional numerical simulations (see Supplementary Material), we obtain the value ${\eta_{\textup{g}}\approx2.2\,\%}$ when the light beam is focused on the edge of a Pt stripe. The spin current reaching the position of the detector is ${i_\mathrm{s,0}\,e^{-x_{0}/L_\mathrm{s}}}$, the exponential term accounting for the spin depolarization along the distance ${x_{0}}$ from the generation point to the detector. Because of the built-in electric field of the junction, only a fraction $\eta_{\textup{t}}$ of the spin-polarized electrons reaching the detector position effectively enters the detector and thus contributes to the measured signal. We calculate $\eta_{\textup{t}}$ by applying the same numerical simulations detailed in Ref.~\citenum{Zucchetti2019}, which account for the height of the Schottky barrier $\Phi_{\textup{bar}}$ between the detector and the Ge substrate and the barrier reduction $\Phi_{\textup{ph}}$ produced by the photovoltaic effect. By performing transport measurements, we obtain the value ${\Phi_{\text{bar}}\approx 0.66~\text{eV}}$ in both cases, as expected from the Fermi level pinning for Ge surfaces \cite{Dimoulas2006}. We also calculate the value ${\Phi_{\text{ph}}=0.29~\text{eV}}$ for both samples using the Nextnano software \cite{Birner2007}. With these parameters, we obtain ${\eta_{\textup{t,BiSe}}}={\eta_{\textup{t,Pt}}=13\,\%}$.

We validate the numerical model with Pt, for which the spin Hall angle has been addressed by several works in the literature. In this case, we have measured ${\Delta V_\mathrm{ISHE}/W=7~\text{nV/\textmu W}}$ (obtained for an incident optical power ${W=15~\text{\textmu W}}$) at ${x=x_0\approx6~\text{\textmu m}}$, corresponding to ${i_{\textup{c}}=\Delta V_\mathrm{ISHE}/R=210~\text{pA}}$. Our numerical estimation of ${i_{\textup{s}}}$ yields ${i_{\textup{s}}=6~\text{nA}}$, giving ${\gamma_{\textup{Pt}}=i_{\textup{c}}/i_{\textup{s}}\approx3.5\,\%}$. This value is comparable to previously reported ones for evaporated Pt films \cite{Sagasta2016,Huo2017}, therefore we apply the same model to the sample with a Bi$_2$Se$_3$ detector. At ${x=x_0\approx6~\text{\textmu m}}$, we obtain ${\Delta V_{\textup{IREE}}/W=-40~\textup{nV/\textmu W}}$ (measured with ${W=18~\text{\textmu W}}$). Hence ${i_{\textup{c}}=-72~\text{pA}}$ and ${i_{\textup{s}}=7.2~\text{nA}}$, yielding ${\gamma_{\textup{BiSe}}\approx-1\,\%}$.

Since the spin-to-charge conversion by the IREE occurs in surface states, the relevant parameter describing the SCC efficiency is the inverse Rashba-Edelstein length $\lambda_{\textup{IREE}}$, which can be obtained as the product between the macroscopic efficiency parameter $\gamma_{\textup{BiSe}}$ and the spatial extension $d$ of the surface states in which the conversion takes place. Taking ${d=3~\textup{nm}}$ from Ref.~\citenum{Neupane2014}, we find ${\lambda_{\textup{IREE}}=\gamma_{\textup{BiSe}}\,d\approx-30~\textup{pm}}$. Note that, to derive $\lambda_{\textup{IREE}}$ from $\gamma_{\textup{BiSe}}$, we only need to consider SCC occurring in the Bi$_2$Se$_3$ surface states in contact with Ge and neglect SCC at the opposite Bi$_2$Se$_3$ free surface, since the film is thicker than the spin diffusion length \cite{Deorani2014}.

\begin{figure}[ht]
\begin{center}
\includegraphics[width=0.48\textwidth]{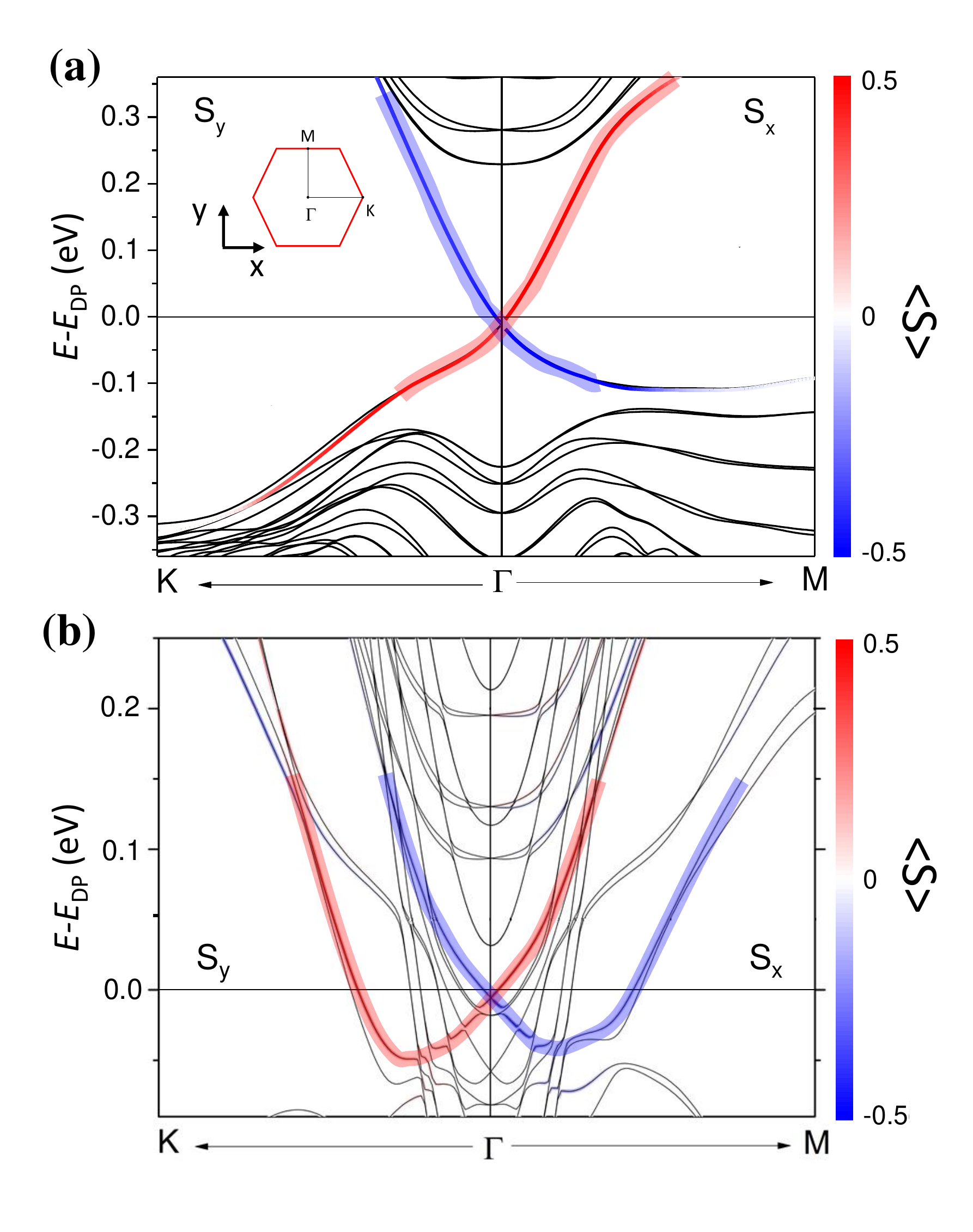}
\end{center}
\caption{(Color online) (a) Electronic band structure with spin-orbit coupling for 8 QL of Bi$_2$Se$_3$. The red and blue colors indicate the spin texture of the topmost QL projected along the $y$ and $x$ direction for the K-$\Gamma$ and $\Gamma$-M high symmetry axes, respectively. The corresponding Brillouin zone is reported in the inset. (b) Electronic band structure for the Bi$_2$Se$_3$ (8 QL)/Ge (3.2 nm) stack. The same color code as in (a) is used for the interface spin texture. $E_{\textup{DP}}$ corresponds to the energy position of the Dirac point of Bi$_2$Se$_3$ surface states in (a) and Bi$_2$Se$_3$/Ge interface states in (b).} \label{fig5}
\end{figure}

In order to understand the opposite SCC signs for the two samples, we have performed first principle relativistic calculations to unveil the spin-resolved band structure at the Bi$_2$Se$_3$/Ge interface (see Supplementary Material). We first consider eight quintuple layers (8 QL) of Bi$_2$Se$_3$. In Fig.~\ref{fig5}\textcolor{blue}{(a)}, the band structure is plotted along the K-$\Gamma$-M direction as shown in the inset. In this particular direction along which K-$\Gamma$ ($\Gamma$-M) is parallel to the $x$ ($y$) direction, we plot the band structure weighted by the $y$ ($x$) spin component $S_y$ ($S_x$) of the topmost QL, as highlighted by the thick red and blue lines. The red (blue) color indicates an in-plane spin pointing in the positive (negative) direction of the axis. We clearly observe the presence of surface states belonging to Dirac cones. Due to spin-momentum locking characteristic of TIs, the in-plane spin helicity of the surface states above the Dirac point (characterized by a positive dispersion) displays a clockwise (CW) chirality, while the helicity of states below the Dirac point (with a negative dispersion) is counterclockwise (CCW). Because of the opposite dispersion relation, both types of chiral states (either above or below the Dirac point) thus lead to a \textit{positive} $\lambda_{\textrm{IREE}}$ value and to the same sign of the SCC coefficient as the one observed in platinum \cite{RojasSanchez2016}.

Figure~\ref{fig5}\textcolor{blue}{(b)} displays the band structure of 8 QL of Bi$_2$Se$_3$ in contact with 3.2~nm of Ge. Compared with pure Bi$_2$Se$_3$, many additional electronic states appear due to the strong hybridization with Ge. In Fig.~\ref{fig5}\textcolor{blue}{(b)}, we use the same color code as in Fig.~\ref{fig5}\textcolor{blue}{(a)} to highlight the spin texture at the Bi$_2$Se$_3$/Ge interface. Interestingly, due to the strong hybridization between Bi$_2$Se$_3$ and Ge orbitals, the bottom Dirac cone is inverted. This cone inversion gives rise to a Rashba-like helical spin texture exhibiting a counter-clockwise (CCW) chirality of the outer contour for -0.05 eV $< E-E_{\textup{DP}}<$ 0.15 eV, $E_{\textup{DP}}$ being the energy of the Dirac point. Therefore, in this energy range, the CCW spin chirality of the outer contour leads to a \textit{negative} $\lambda_{\textrm{IREE}}$ value. First principles calculations thus qualitatively support our experimental observations concerning the sign of the spin-charge conversion. It should also be noticed that, by adjusting the position of the Fermi level in Fig.~\ref{fig5}\textcolor{blue}{(b)} with a gate voltage to the Bi$_2$Se$_3$/Ge heterostructure, it could be possible to control both the magnitude and the sign of the spin-to-charge conversion at the interface.

To summarize, we have probed the spin-to-charge conversion at the Bi$_2$Se$_3$/Ge interface by using a non-local spin injection/detection scheme. The study of a comparison sample that only differs by the spin detection mechanism (ISHE in Pt), allows gaining insight into the spin-to-charge conversion mechanisms taking place at the TI/semiconductor interface. 
Notably, we measure larger voltage drops with Bi$_2$Se$_3$ than with Pt, which makes the former an excellent spin detector for future spin-based technologies. We have numerically modeled the spin injection and transport in Ge to the detector and retrieved a spin Hall angle for Pt comparable with values previously reported in literature. The same model applied to Bi$_2$Se$_3$/Ge yields an equivalent spin-Hall angle close to one derived for Pt and corresponding to an inverse Rashba-Edelstein length ${\lambda_{\textup{IREE}}\approx-30~\text{pm}}$. The sign of the spin-to-charge conversion is found to be opposite for Bi$_2$Se$_3$/Ge and Pt. By employing first principles calculations, we ascribe this behavior to the interfacial hybridization between the topologically protected surface states of Bi$_2$Se$_3$ and Ge leading to the formation of Rashba interface states with a spin chirality opposite to the one of states at the free Bi$_2$Se$_3$ surface. Our results demonstrate that semiconductors constitute a very promising platform for the exploitation of topological insulators in spintronics, where, by gating the heterostructure, spin-to-charge conversion could in principle be tuned in magnitude and sign.\\
The authors acknowledge the financial support from the ANR project ANR-16-CE24-0017 TOP RISE and from the European Union’s Horizon 2020 research and innovation programme under Grant agreement No. 785219 (Graphene Flagship).

\end{sloppypar}

\end{document}